# Tencent Video Dataset (TVD): A Video Dataset for Learning-based Visual Data Compression and Analysis

Xiaozhong Xu, Shan Liu and Zeqiang Li

## ABSTRACT

Learning-based visual data compression and analysis have attracted great interest from both academia and industry recently. More training as well as testing datasets, especially good quality video datasets are highly desirable for related research and standardization activities. Tencent Video Dataset (TVD) is established to serve various purposes such as training neural network-based coding tools and testing machine vision tasks including object detection and tracking. TVD contains 86 video sequences with a variety of content coverage. Each video sequence consists of 65 frames at 4K (3840x2160) spatial resolution. In this paper, the details of this dataset, as well as its performance when compressed by VVC and HEVC video codecs, are introduced.

*Index Terms: 4K Dataset, Video Compression, Machine Learning, Video Coding for Machines, Object Detection, Object Tracking*

## 1. Introduction

In recent years, new machine learning technologies have brought significant improvements on visual data compression and analysis. In these areas, deep neural networks (DNN) or in general learning-based methods enabled advanced compression efficiency and machine task accuracy through a data driven approach.

Efficient learning-based methods are typically designed by going through an extensive training process. Good quality video datasets are therefore highly desirable for related research and standardization activities. JPEG AI, JVET Neural Network based Video Coding (NNVC) and MPEG Video Coding for Machines (VCM) are just a few examples. Tencent Video Dataset (TVD) [1] is established to serve various purposes such as training neural network-based coding tools and testing machine vision tasks including object detection and tracking.

TVD contains 86 video sequences with a variety of content coverage. Each video sequence consists of 65 frames at 3840x2160 spatial resolution. This video dataset has been used as a training set in JVET NNVC exploration activities [2].

For object detection exploration, 166 images are sampled from TVD at 1920x1080 spatial resolution in rgb24 png format. Bounding box annotations for these images are provided. These images with annotation have been included in the common test conditions of MPEG VCM as a testing set [3].

## 2. Video Data collection

All the sequences in TVD are captured using Red Helium 8k, Red Monstro 8K and Blackmagic URSA Mini Pro 12K. The sequences are transcoded and then converted to YUV 4:2:0 color format using FFmpeg [4]. Higher resolution format of the same content is possible as the some of the source video clips are captured in a higher resolution (at least 8K).

A variety of scenes with static or moving objects are included in this dataset. The thumbnails of the entire video dataset are provided on the TVD website [1]. In Fig. 1, some sample frames of the sequences in the TVD are presented.

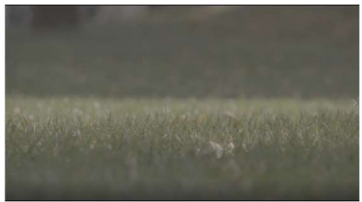
Grass swinging

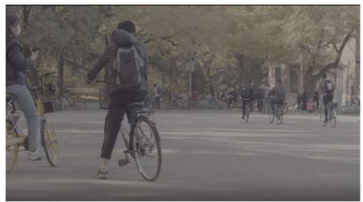
Bikes passing by

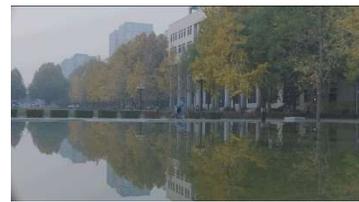
Static water with bikes passing by

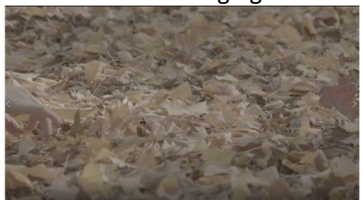
Fallen leaves on the ground

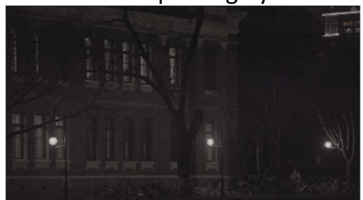
Building in the dark scene

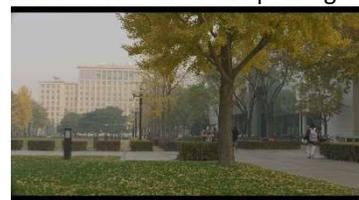
Trees with bikes passing by

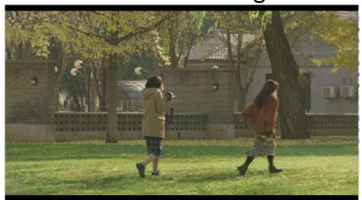
Two girls running on the grass

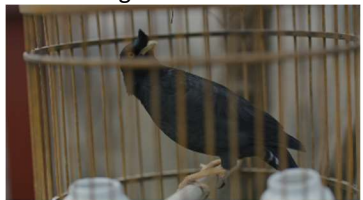
Black bird in the cage

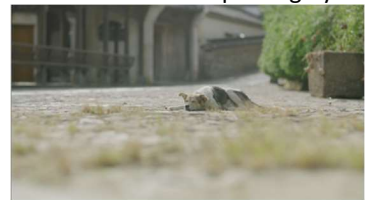
Gog lying on the ground

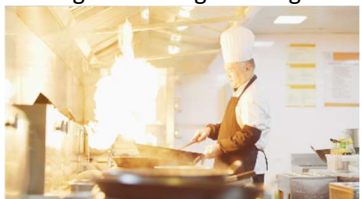
Chef cooking in the kitchen

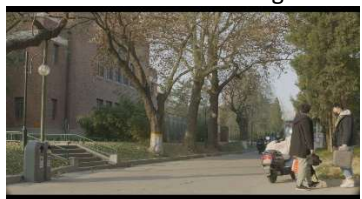
Boys standing on the street

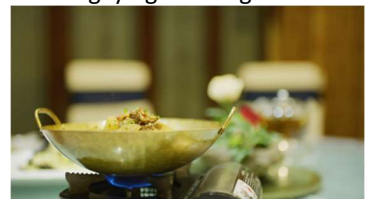
Hot dishes with table spinning

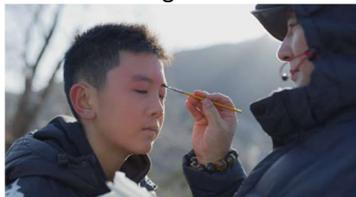
Boy making up on the Great Wall

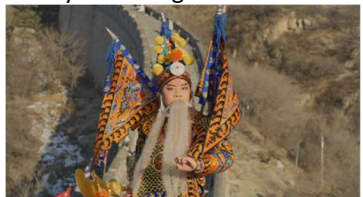
Boy wearing opera costume standing on the Great Wall

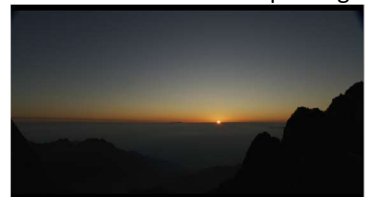
Sunrise at Mountain Huang

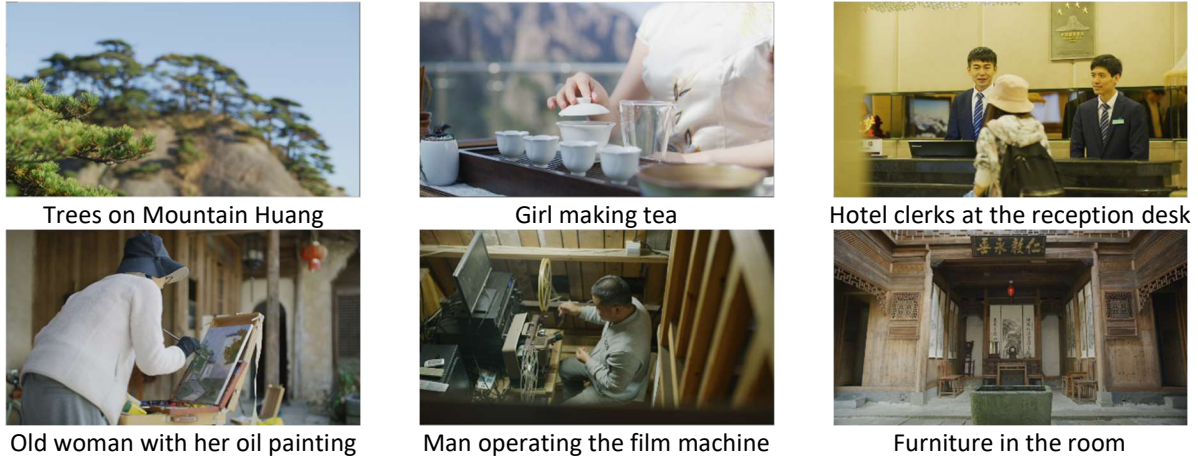

Fig. 1 Some sample frames of 15 example sequences from the TVD

## 3. Evaluation on Video Compression Tools

The behavior of a video sequence is of interest when it is compressed by a typical video codec. In this paper, video encoding and decoding on the TVD using VVC reference software version VTM-11.0 [5] and HEVC reference software version HM 16.23 [6] are performed. The overall result statistics as well as some selected sequences are reported.

### 3.1 Compression results and comparison

With configurations of AI (All intra)/RA (Random Access)/LDB (Low Delay B)/LDP (Low Delay P), and QP equals to {22, 27, 32, 37, 42}, VTM-11.0 and HM-16.23 are launched for video encoding/decoding on the TVD, using the default encoding parameters in the software packages. In the following tables, the average statistics of the entire dataset are presented. For HM-16.23, the average of PSNR for Y/U/V channel for different QPs is showed in Table I and II. For VTM-11.0, the average of PSNR for Y/U/V channel for different QPs is showed in Table III and IV.

Table I: Average PSNR for YUV channel for different QP with HM for TVD (AI/RA)

| QP | AI | | | | RA | | | |
|----|---------|--------|--------|--------|---------|--------|--------|--------|
|    | Bitrate | Y-PSNR | U-PSNR | V-PSNR | Bitrate | Y-PSNR | U-PSNR | V-PSNR |
| 22 | 105391.43 | 48.88 | 48.14 | 48.69 | 9157.87 | 48.55 | 48.15 | 48.80 |
| 27 | 56468.74 | 46.44 | 46.30 | 47.00 | 4068.81 | 46.47 | 46.78 | 47.50 |
| 32 | 30942.50 | 43.85 | 44.79 | 45.63 | 2132.87 | 44.06 | 45.24 | 46.07 |
| 37 | 16821.47 | 41.15 | 43.63 | 44.57 | 1134.81 | 41.44 | 43.84 | 44.80 |
| 42 | 8702.63 | 38.40 | 42.42 | 43.46 | 609.31 | 38.75 | 42.99 | 44.01 |

Table II: Average PSNR for YUV channel for different QP with HM for TVD (LDB/LDP)

| QP | LDB | | | | LDP | | | |
|----|---------|--------|--------|--------|----------|--------|--------|--------|
|    | Bitrate | Y-PSNR | U-PSNR | V-PSNR | Bitrate  | Y-PSNR | U-PSNR | V-PSNR |
| 22 | 9797.13 | 47.92 | 47.56 | 48.24 | 10474.01 | 47.71 | 47.48 | 48.17 |
| 27 | 3377.45 | 45.48 | 46.01 | 46.82 | 3566.34 | 45.27 | 45.94 | 46.74 |
| 32 | 1633.11 | 42.90 | 44.71 | 45.62 | 1707.73 | 42.71 | 44.62 | 45.54 |
| 37 | 848.66 | 40.28 | 43.35 | 44.38 | 881.92 | 40.10 | 43.26 | 44.28 |
| 42 | 446.49 | 37.54 | 42.50 | 43.59 | 460.12 | 37.40 | 42.41 | 43.50 |

Table III: Average PSNR for YUV channel for different QP with VTM for TVD (AI/RA)

| QP | AI | | | | RA | | | |
|---|---|---|---|---|---|---|---|---|
| | Bitrate | Y-PSNR | U-PSNR | V-PSNR | Bitrate | Y-PSNR | U-PSNR | V-PSNR |
| 22 | 97744.55 | 49.49 | 48.55 | 49.16 | 8636.60 | 49.15 | 48.65 | 49.36 |
| 27 | 48999.22 | 47.02 | 46.70 | 47.52 | 3649.07 | 47.27 | 47.38 | 48.20 |
| 32 | 26098.24 | 44.54 | 45.05 | 46.03 | 1832.88 | 45.00 | 46.01 | 46.96 |
| 37 | 14028.37 | 41.91 | 43.73 | 44.64 | 958.89 | 42.54 | 44.37 | 45.24 |
| 42 | 7226.64 | 39.23 | 42.47 | 43.50 | 503.15 | 39.96 | 43.17 | 44.14 |

Table IV: Average PSNR for YUV channel for different QP with VTM for TVD (LDB/LDP)

| QP | LDB | | | | LDP | | | |
|---|---|---|---|---|---|---|---|---|
| | Bitrate | Y-PSNR | U-PSNR | V-PSNR | Bitrate | Y-PSNR | U-PSNR | V-PSNR |
| 22 | 8815.02 | 48.27 | 47.87 | 48.63 | 9133.07 | 48.21 | 47.84 | 48.59 |
| 27 | 2776.54 | 45.96 | 46.49 | 47.37 | 2860.55 | 45.91 | 46.46 | 47.34 |
| 32 | 1307.67 | 43.49 | 45.00 | 46.01 | 1334.70 | 43.45 | 44.97 | 45.97 |
| 37 | 671.52 | 40.94 | 43.55 | 44.50 | 678.89 | 40.91 | 43.52 | 44.47 |
| 42 | 341.01 | 38.31 | 42.32 | 43.39 | 343.80 | 38.28 | 42.29 | 43.36 |

## 3.2 Comparison between VTM-11.0 and HM-16.23

The overall BD-Rate [7] performance between VTM-11.0 and HM-16.23 are compared, as shown in Table V. Tt is demonstrated that VTM-11.0 provide better BD-Rate result over HM-16.23 among all YUV channels for the TVD. Roughly about 33% BD-rate reduction under RA configuration can be achieved using VTM reference software, when compared with the HM software. The distortion metric in these tests is selected as PSNR.

Table V: BD-Rate gain of VTM-11.0 over HM-16.23 on TVD

| Testing Configuration | BD-rate changes (VTM-11.0 over HM-16.23) | | | | |
|---|---|---|---|---|---|
| | Y-PSNR | U-PSNR | V-PSNR | EncT | DecT |
| All Intra | -27.77% | -24.65% | -25.96% | 557% | 173% |
| Random Access | -32.97% | -35.95% | -37.72% | 210% | 144% |
| Low Delay B | -30.50% | -31.14% | -32.39% | 203% | 135% |
| Low Delay P | -34.52% | -34.38% | -36.98% | 199% | 137% |

Showing the RD curves of the entire video dataset requires too much work and may not be as informative as expected. As showed in Fig. 2 to Fig. 5, the RD curves of a set of selected sequences in the TVD are presented to illustrate the performance result comparison between VTM-11.0 and HM-16.23 for video encoding/decoding on the TVD, with configuration of AI (All intra)/RA (Random Access)/LDB (Low Delay B)/LDP (Low Delay P), respectively. For sequences of BoyWithCostume, BuildingTouristAttraction, GirlRunningOnGrass and TreeAndLeaves, they are considered to be among the sequences with highest bitrates when compressed with the same QP. On the other hand, sequences of ChefCooking5, FilmMachine, SunriseMountainHuang, GrassLand are among the sequences with lowest bitrates when compressed with the same QP.

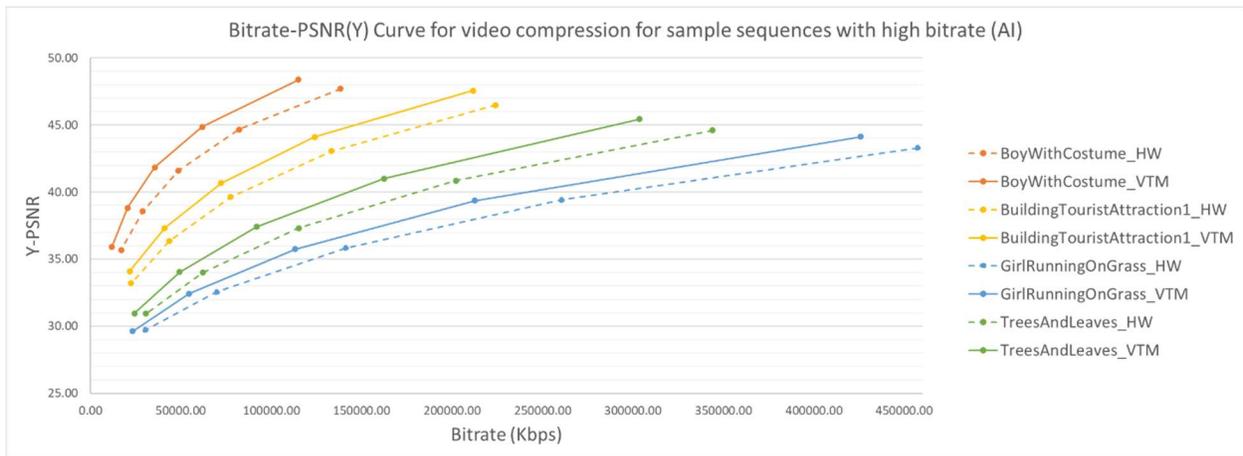

Fig. 2.1 RD curves comparison between VTM-11.0 and HM-16.23 for high bitrate sequences (AI)

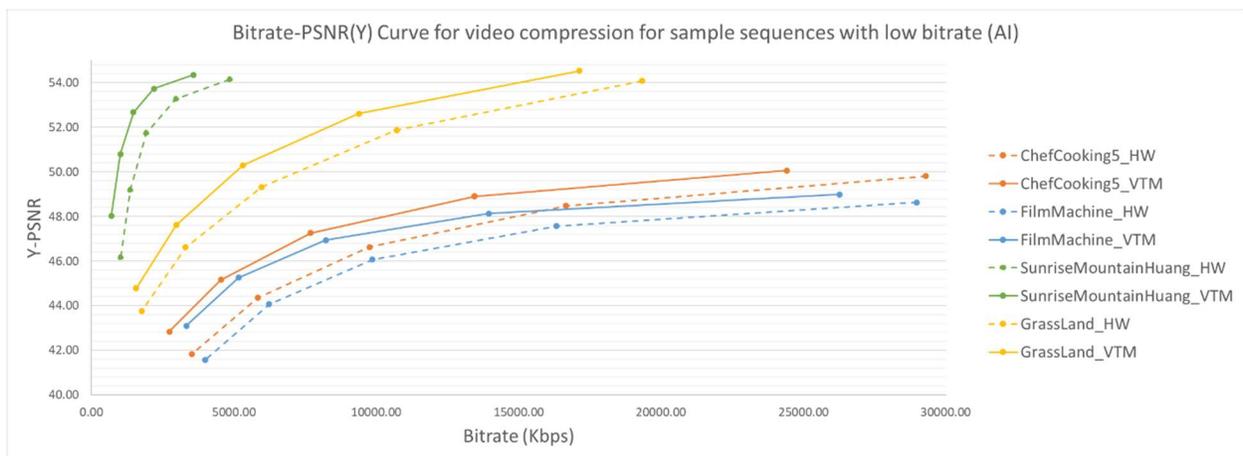

Fig. 2.2 RD curves comparison between VTM-11.0 and HM-16.23 for low bitrate sequences (AI)

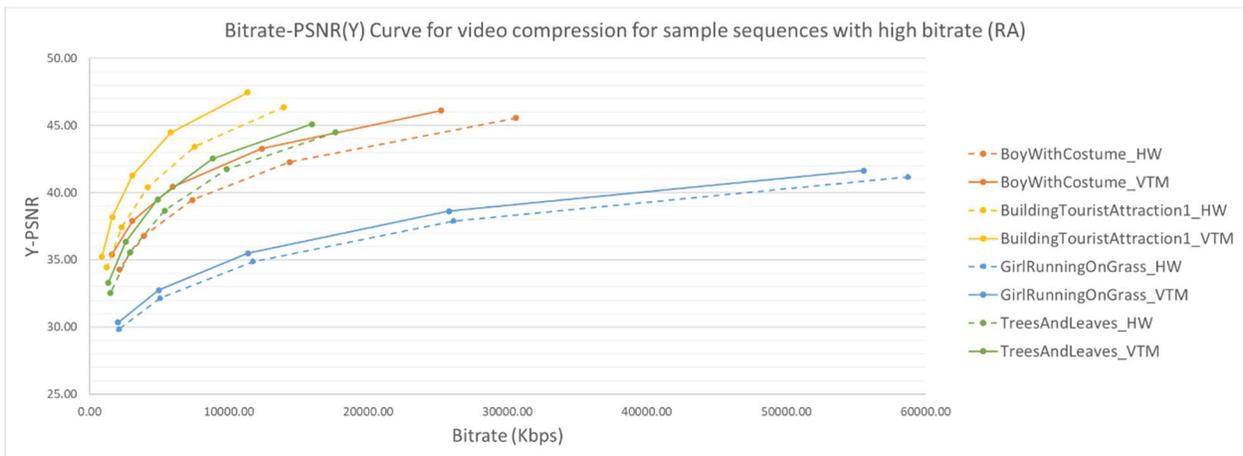

Fig. 3.1 RD curves comparison between VTM-11.0 and HM-16.23 for high bitrate sequences (RA)

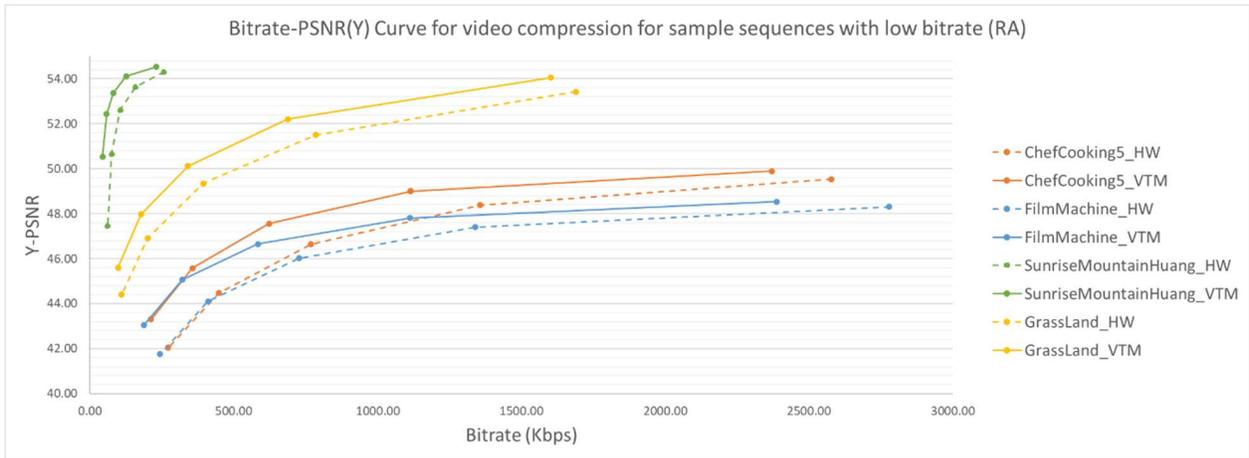

Fig. 3.2 RD curves comparison between VTM-11.0 and HM-16.23 for low bitrate sequences (RA)

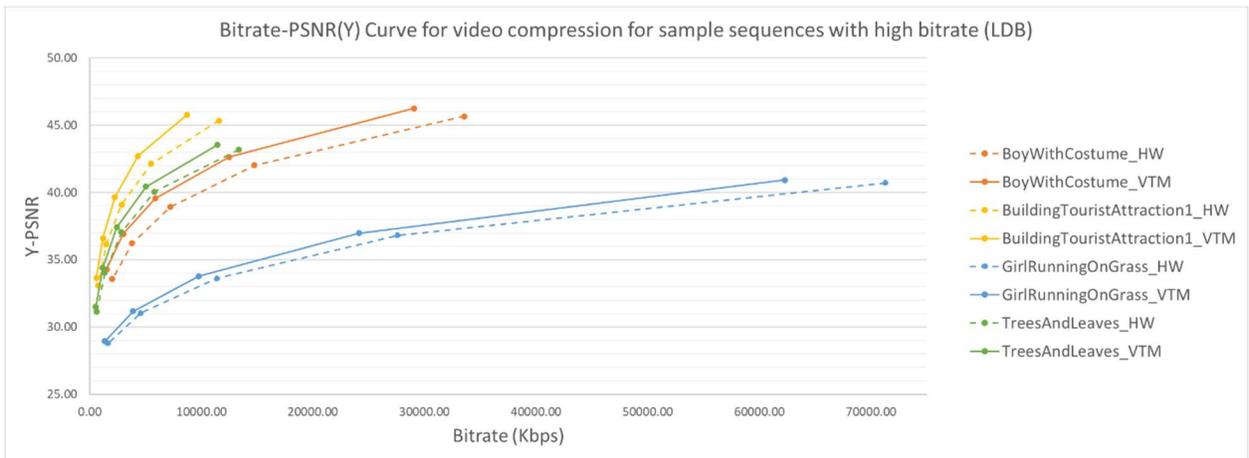

Fig. 4.1 RD curves comparison between VTM-11.0 and HM-16.23 for high bitrate sequences (LDB)

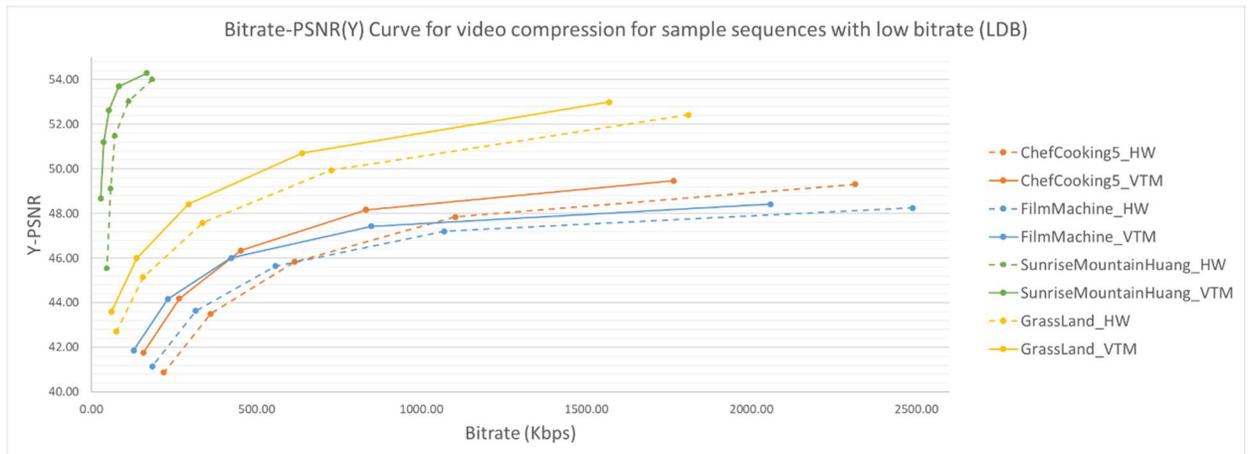

Fig. 4.2 RD curves comparison between VTM-11.0 and HM-16.23 for low bitrate sequences (LDB)

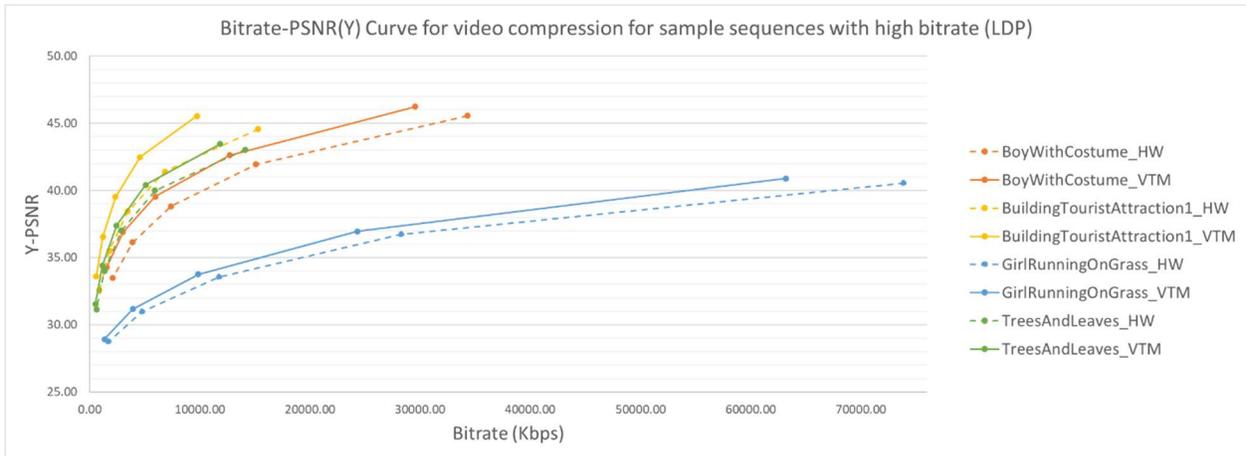

Fig. 5.1 RD curves comparison between VTM-11.0 and HM-16.23 for high bitrate sequences (LDP)

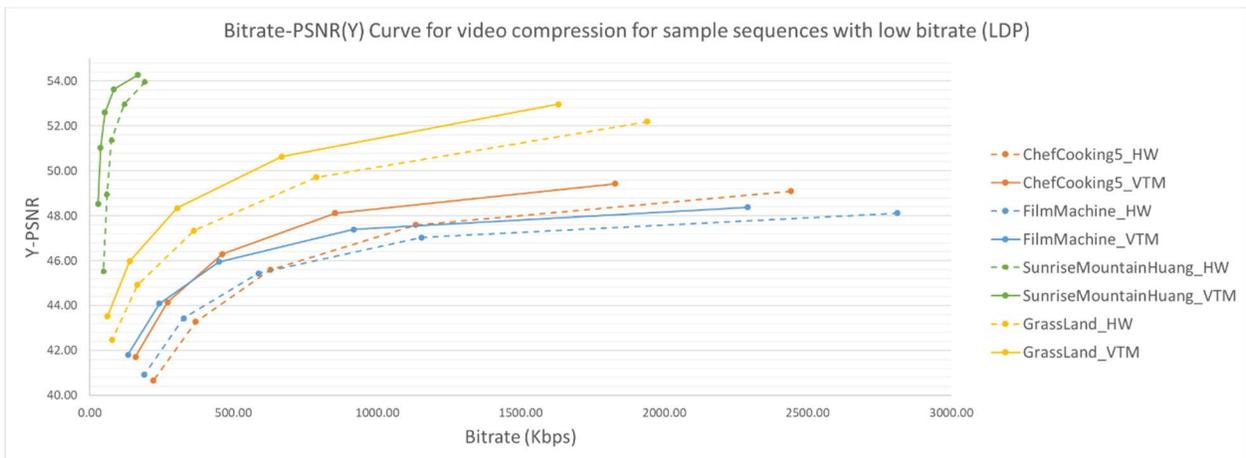

Fig. 5.2 RD curves comparison between VTM-11.0 and HM-16.23 for low bitrate sequences (LDP)

From the above figures, the most difficult (high bitrate) sequences operate at as high as 60 Mbps for RA configuration when compressed with QP22; on the other hand, the set of simple (low bitrate) sequences can be compressed at below 3 Mbps. The dramatic differences in bitrate represent the diversity of the contents in this dataset. From Table I~IV, on average, the high-end operating bitrate is around 8~9 Mbps, which are considered as reasonable for the 4K contents.

## 4. Copyright information

The proposed sequences are captured and produced by Tencent, Shenzhen Boyan Technology Ltd. and Tsinghua University. All intellectual property rights remain with Tencent, Shenzhen Boyan Technology Ltd. and Tsinghua University.

The following uses are allowed for the contributed sequences:

1. Sequences may be published in technical papers, played at technology research and development events.
2. Sequences may be used by standards activities. (e.g., ITU, MPEG, VQEG).
3. The following uses are NOT allowed for the contributed sequences:
4. Do not publish snapshots in product brochures.
5. Do not use video for marketing purposes.
6. Redistribution is not permitted.
7. Do not use in television shows, commercials, or movies.

## 5. Conclusion

This paper presents a new UHD dataset TVD. Diverse contents are included in this dataset with 4K resolution. It has been evaluated by two different video codecs (VVC and HEVC) with four coding configurations. This dataset fits in various of purposes in machine learning related research and standardization activities. Specifically, it has been used in training in NN-based video compression algorithms. With proper annotations, this dataset is further utilized as a test set in video analysis and machine-oriented tasks such as object detection, object segmentation and object tracking.